\documentclass[aps,prb,twocolumn,superscriptaddress,showpacs]{revtex4}

\usepackage{color}
\usepackage{graphicx}

\begin{document}
%\preprint{}
\input epsf.sty
\title{Synchrotron X-ray diffraction study of a charge stripe order in $1/8$-doped 
La$_{1.875}$Ba$_{0.125-x}$Sr$_{x}$CuO$_{4}$}


\author{H. Kimura}
\email[]{kimura@tagen.tohoku.ac.jp}
%\thanks{}
\affiliation{Institute of Multidisciplinary Research for Advanced Materials, 
Tohoku University, Sendai 980-8577, Japan}

\author{H. Goka}
\affiliation{Institute for Chemical Research, Kyoto University, Gokasho, Uji 610-0011, Japan}

\author{M. Fujita}
\affiliation{Institute for Chemical Research, Kyoto University, Gokasho, Uji 610-0011, Japan}

\author{Y. Noda}
\affiliation{Institute of Multidisciplinary Research for Advanced Materials, 
Tohoku University, Sendai 980-8577, Japan}

\author{K. Yamada}
\affiliation{Institute for Chemical Research, Kyoto University, Gokasho, Uji 610-0011, Japan}

\author{N. Ikeda}
\affiliation{Japan Synchrotron Radiation Research Institute, Hyogo 679-5198, Japan}


\date{\today}

\begin{abstract}
Lattice distortions associated with charge stripe order in 1/8 hole-doped 
La$_{1.875}$Ba$_{0.125-x}$Sr$_{x}$CuO$_{4}$ are studied using synchrotron 
X-ray diffraction for $x=0.05$ and  $x=0.075$. The propagation wave vector and 
charge order correlation lengths are determined with a high accuracy, 
revealing that the oblique charge stripes in orthorhombic $x=0.075$ 
crystal are more disordered than the aligned stripes in tetragonal $x=0.05$ crystal.
The twofold periodicity of lattice modulations along the $c$-axis is explained by long-range 
Coulomb interactions between holes on neighboring CuO$_{2}$ planes.
\end{abstract}

%\keywords{}

\maketitle

%\section{Introduction}
The interplay between spin and charge correlations in hole-doped CuO$_{2}$ planes is widely 
believed to be related to the mechanisms of high-$T_{\rm c}$ superconductivity.
In La$_{2-x}$Ba$_{x}$CuO$_{4}$, which is a prototypical high-$T_{\rm c}$ superconductor, 
anomalous suppression of superconductivity has been observed at around a specific hole 
concentration of $x=1/8$, where the Low-Temperature-Tetragonal (LTT) crystal phase 
($P4_{2}/ncm$ symmetry) occurs\cite{Moodenbaugh1988,Axe1989}.
Tranquada {\it et al}. have found the incommensurate spin- and charge orders 
in the LTT phase of La$_{1.6-x}$Nd$_{0.4}$Sr$_{x}$CuO$_{4}$ (LNSCO) with $x=0.12$
\cite{Tranquada1995,Tranquada1996,Tranquada1997}.
The results revealed that a strong relation exists between spin/charge ordering, crystal 
structure, and the suppression of high-$T_{\rm c}$ superconductivity. 
Based on the stripe model\cite{Tranquada1995,Kivelson1998}, these relationships can 
be explained by the pinning of dynamical 
charge stripe correlations by lattice potentials, 
resulting in the strong suppression of superconductivity.
Recently, a systematic neutron scattering study of the incommensurate spin/charge order 
in La$_{1.875}$Ba$_{0.125-x}$Sr$_{x}$CuO$_{4}$ (LBSCO) with $0.05\leq x \leq 0.085$ 
has confirmed that charge ordering only occurs in LTT and LTLO 
(Low-Temperature-Less-Orthorhombic, {\it Pccn} symmetry) phases and competes with 
superconductivity, whereas the robustness of magnetic order depends weakly on crystal 
structure and $T_{\rm c}$ suppression compared to charge order\cite{Fujita2002}. 
Hence, an understanding of the microscopic nature of charge order is important for 
clarifying the relationship between charge correlation and superconductivity. 

Although charge order is observed as lattice distortions in neutron scattering, X-ray 
diffraction can, in principle, directly detect charge distributions, which would provide 
direct evidence of charge order. 
A recent synchrotron X-ray diffraction study of LNSCO at $x=0.12$ 
has determined the propagation wave vector of the incommensurate charge order, 
$Q_{\rm ch}=(\pm 2\epsilon\ 0\ \frac{1}{2})$ with 
$\epsilon=0.118$~r.l.u. (reciprocal lattice unit)\cite{Zimmermann1998}. 
Although the superlattice observed in the X-ray diffraction study was mainly the result of 
lattice distortions, precise determination of the wave vector $Q_{\rm ch}$ revealed that 
the lattice distortions are caused by the formation of charge stripe order.
In LBSCO systems, neutron scattering measurements have found that the in-plane component 
of $Q_{\rm ch}$ for $x=0.05$ in the LTT structure is different from that for $x=0.075$ in 
the LTLO structure, which suggests a strong relationship between stripe pattern and 
crystal symmetry\cite{Fujita2002_2}.
However, detailed information about the three dimensional correlation of the charge 
order is not available yet because no synchrotron X-ray diffraction measurements 
have been carried out in LBSCO systems.

Synchrotron X-ray diffraction measurements of LBSCO with $x=0.05$ and 0.075 are 
conducted to study the nature of charge stripe order in detail, and to examine the 
relationship between charge correlation and crystal structure.
%\section{Experimental}

X-ray diffraction experiments were performed at the Crystal Structure Analysis Beam 
Line (BL02B1)\cite{Noda1998} of SPring-8.
X-ray energy was tuned to 30~keV using a sagittally bent Si(311) double monochromator. 
A double platinum mirror vertically collimates the incident beam and completely eliminates 
higher order harmonics. Single crystals of LBSCO with $x=0.05$ and $x=0.075$ were obtained 
from the same batch as crystals used in previous neutron scattering 
studies\cite{Fujita2002,Fujita2002_2,Fujita2001}. 
The cylindrical crystals are about 5~mm in diameter with a height of 1~mm. 
The reciprocal lattice is defined in the $I{\rm 4}/mmm$ symmetry where the two 
short axes correspond to the distance between the nearest-neighbor Cu atoms along 
the in-plane Cu-O bond. At the (2~0~0) point, the longitudinal resolution 
($ \Vert a^{*} $-axis) was about 0.014~\AA$^{-1}$ 
and transverse resolutions along the $b^{*}$- and the $c^{*}$- axes 
were $\sim 0.005$~\AA$^{-1}$ and $\sim 0.046$~\AA$^{-1}$, respectively.

\begin{figure}[t]
\centerline{\epsfxsize=2.34in\epsfbox{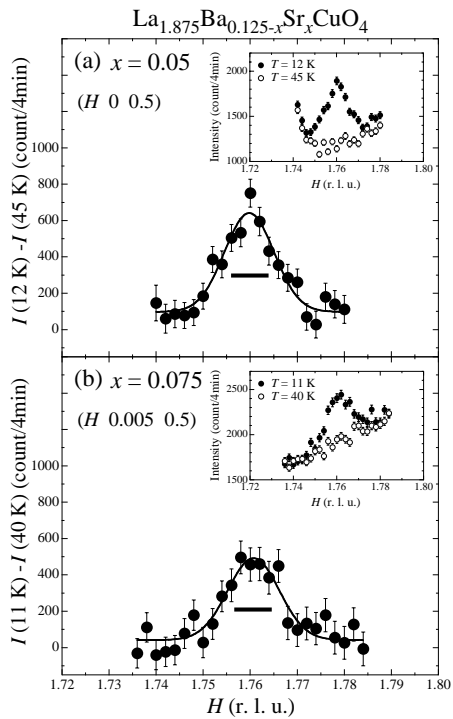}}
\caption{$H$-scan profiles of superlattice peaks for (a) $x=0.05$ and (b) $x=0.075$ crystals. 
Backgrounds have been subtracted. The insets shows the raw data taken at around 
12~K (closed circle) and around 45~K (open circle). 
Bold horizontal lines indicate instrument resolution.}
\label{fig1}
\end{figure}

%\section{Results}
Figures~\ref{fig1}(a) and (b) show $H$-scan profiles of superlattice peaks around 
$Q=(2-2\epsilon\ 0\ 0.5)$ for $x=0.05$ and $x=0.075$, respectively, after 
%=========================================================
%=========================================================
adjusting against a background measurement. Raw data is shown in the inset of each figure.
As can be seen in the figures, the superlattice peaks for both $x=0.05$ and $x=0.075$ 
crystals are clearly located at $H=1.76$~r.l.u.
Thus the incommensurability $2\epsilon$ is exactly $0.240\pm 0.001$~r.l.u.
In a previous neutron scattering study\cite{Fujita2001}, the incommensurability of 
elastic magnetic peaks of $x=0.05$ samples was found to be $\epsilon=0.120\pm 0.001$~r.l.u., 
indicating that the superlattice peaks observed in the present study indeed correspond to 
second-order harmonics of magnetic order.
The line-widths are clearly broadened with respect to instrument resolution (indicated by 
bold horizontal lines in the figures), which gives finite in-plane correlation lengths along the 
$a$-axis ($\equiv\xi_{a}$) of $130\pm 20$~\AA\ and $120\pm 30$~\AA\ for 
$x=0.05$ and $x=0.075$, respectively. 

\begin{figure}[t]
\centerline{\epsfxsize=2.34in\epsfbox{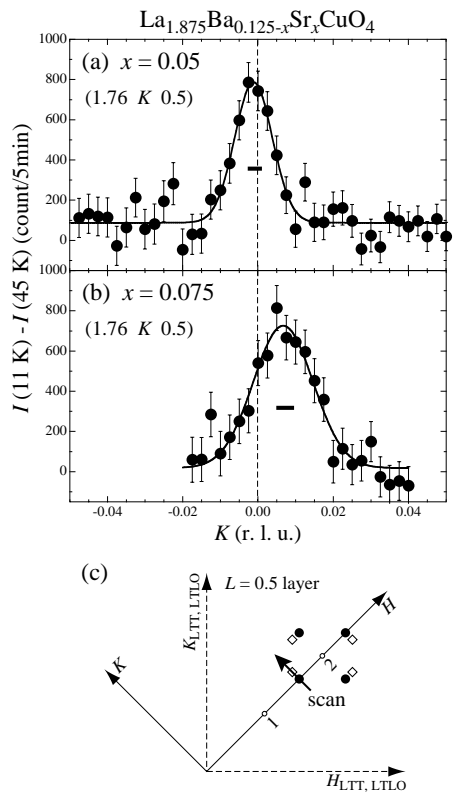}}
\caption{$Q$-profiles of superlattice peaks along $K$-direction for (a) $x=0.05$ and (b) $x=0.075$ crystals. 
Backgrounds measured at 45~K have been subtracted. 
Bold horizontal lines indicate instrument resolution. A trajectory of the $q$-scan 
and schematic peak positions of the superlattice are shown in (c).
Solid circles and open diamonds correspond to the positions for $x=0.05$ and 0.075, respectively.}
\label{fig2}
\end{figure}

$K$-scan profiles of superlattice peaks at around $Q=(2-2\epsilon\ 0\ 0.5)$ 
are shown in Figs.~\ref{fig2}(a) and (b) after background correction. 
%=========================================================
%========================================================= 
Figure~\ref{fig2}(c) shows a trajectory of the $q$-scan and the locations 
of superlattice peaks in reciprocal lattice space. 
Note that the $K$-direction is perpendicular to the propagation 
wave vector $Q_{\rm ch}$. The peak for the $x=0.05$ crystal is almost at $K=0$ 
(indicated by a dashed line in the figure), whereas for the $x=0.075$ crystal, the peak is clearly 
shifted away from $K=0$. The amplitude of the peak shift was found to be 
$0.007\pm 0.001$~r.l.u, the same as found in a previous neutron scattering study\cite{Fujita2002_2}.
In addition, the $x=0.075$ crystal used in the x-ray diffraction was composed of a single 
domain unlike the crystal from the neutron scattering study\cite{Fujita2002_2}, 
which contained a twin due to the orthorhombic symmetry of {\it Pccn}. 
Hence, the shift of the superlattice peak in the $x=0.075$ crystal 
is clearly not an artifact, with the quartet of superlattice peaks 
forming a regular rectangular shape in reciprocal lattice space, 
as shown by the open diamonds in Fig.~\ref{fig2}(c). 
This arrangement of peaks satisfies the orthorhombic symmetry of the 
LTLO phase, {\em not} the tetragonal symmetry of the LTT phase, indicating that 
the pattern of charge order is closely related to crystal structure.
The in-plane correlation length along the $b$-axis ($\equiv\xi_{b}$) for the tetragonal 
$x=0.05$ crystal was $110\pm 10$~\AA, and similar for $\xi_{a}$.
In comparison, $\xi_{b}$ of the orthorhombic $x=0.075$ ($=70\pm 8$~\AA) is 
clearly shorter than $\xi_{a}$ and also shorter than the $\xi_{b}$ of the $x=0.05$ crystal. 

\begin{figure}[t]
\centerline{\epsfxsize=2.34in\epsfbox{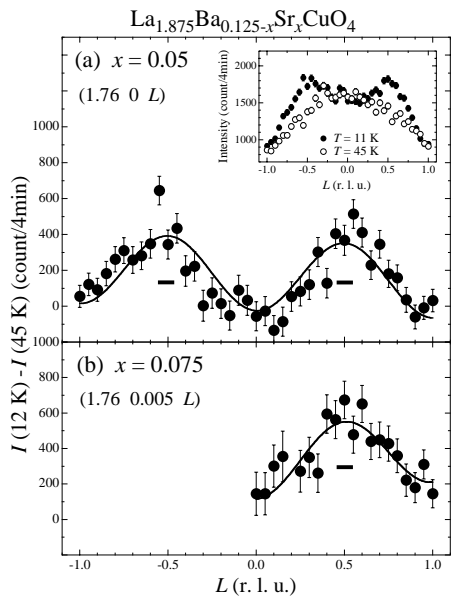}}
\caption{$L$-scan profiles of superlattice peaks for (a) $x=0.05$ and (b) $x=0.075$ crystals.
Inset in (a) shows raw data taken at 12~K (closed circle) and 45~K (open circle). 
The short horizontal line indicates instrument resolution.}
\label{fig3}
\end{figure}

Figure~\ref{fig3} shows the $L$-dependence of the superlattice peaks around 
%=========================================================
%=========================================================
$Q=(2-2\epsilon\ \ 0\ \pm 0.5)$ for (a) $x=0.05$ and (b) $x=0.075$, corresponding to 
out-of-plane correlations. The plots show the difference between data at $T=11$~K and 45~K.
Raw data for the $x=0.05$ crystal at each temperature is plotted in the inset of 
Fig.~\ref{fig3}(a). Intensities of both the samples modulate sinusoidally and exhibit 
broad maxima at $L=\pm 0.5$~r.l.u., indicative of a twofold periodicity along $c$-axis. 
The line-width is much broader than instrument resolution. Thus a reasonably short 
out-of-plane correlation length $\xi_{c}$ of 
$\sim 9$~\AA\ was obtained for both the $x=0.05$ and 0.075 crystals, which is 
shorter than that the next-nearest-neighbor (n.n.n.) distance between CuO$_{2}$ planes. 
The large anisotropy between $\xi_{a,b}$ and $\xi_{c}$ suggests two-dimensional 
charge correlations. Solid curves in Fig.~\ref{fig3} denote fits to the equation 
$|F(L)|^{2}\propto |1-e^{-i2\pi L}|^{2}=4\sin(\pi L)$. 
The good agreement of this equation with the data indicates that there is an 
antiphase relationship between n.n.n. CuO$_{2}$ layers, which are separated by a 
distance $c$, which can be explained by 
a long-range Coulomb interaction between doped holes on the CuO$_{2}$ planes. 
The integrated intensity along $L$ of the superlattice peak is $\sim 10^{7}$ times weaker 
than that of the fundamental (2\ 0\ 0) Bragg reflection of intensity $\sim10^{8}$ cps. 
In addition, the relative intensity of superlattice peak to the fundamental peak is $\sim 10$ 
times weaker than found in the neutron scattering study. 
These results show that lattice distortions are the main contributor to the superlattice 
intensity and that the relative intensity is qualitatively consistent with a model in which 
the largest atomic displacement resulting from charge order is oxygen.
The amplitude of oxygen displacement along the $a$-axis can be estimated to be less 
than 10$^{-3}$~\AA\ by a simple calculation based on the stripe model and using the 
measured relative intensities.

\begin{figure}[t]
\centerline{\epsfxsize=2.34in\epsfbox{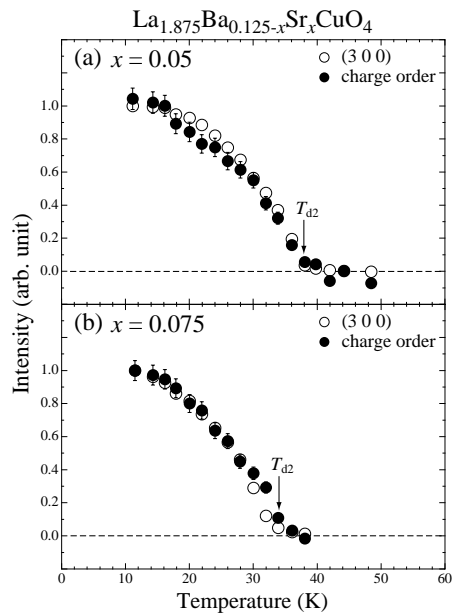}}
\caption{Temperature dependences of (3\ 0\ 0) (open circle) and charge order (closed circle) peaks for (a) $x=0.05$ and (b) $x=0.075$ crystals.}
\label{fig4}
\end{figure}

The temperature dependence of superlattice peak intensity and of the (3\ 0\ 0) reflection, 
which corresponds to the order parameter of structural phase transition into the LTT or 
LTLO phase, were measured. Results are shown in Figs.~\ref{fig4} (a) and (b) for the 
$x=0.05$ and $x=0.075$ crystals, respectively.
Structural phase transition temperatures, $T_{\rm d2}$, are indicated in the figures at the 
%=========================================================
%=========================================================
point where the (3\ 0\ 0) superlattice intensity diminishes. 
The $T_{\rm d2}$ transition temperatures for the $x=0.05$ and $x=0.075$ crystals were 
thus estimated to be 38~K and 34~K, respectively, almost identical to those obtained by 
neutron scattering\cite{Fujita2002}. Remarkably, the temperature dependence of the 
superlattice peak intensity (closed circles) is almost identical to that of the order 
parameter for the LTT/LTLO phase (open circles), suggesting that the ordering process of 
charge order is closely related to that of LTT/LTLO structural phase transition. 
These results are quite a contrast to the LNSCO system, where the 
superlattice peak evolves gradually as temperature decreases whereas 
the LTT order parameter exhibits a first order phase 
transition\cite{Tranquada1996,Zimmermann1998}.

%\section{Discussion}
High-$Q$ resolution as well as the high-statistics of the present X-ray diffraction study have 
provided precise propagation wave vectors of the superlattice peaks associated with charge 
order, giving $Q_{\rm ch} = (\pm 0.24\ \mp\eta\ \frac{1}{2})$ with $\eta=0$ and 
0.007~r.l.u. for $x=0.05$ and $x=0.075$ crystal, respectively.
It is remarkable that the incommensurability $\epsilon=0.12$~r.l.u. of both samples is 
almost identical to that of LNSCO for $x=0.12$ but is inconsistent with hole-doping 
$x=1/8$ of the present samples. 
As can be seen in Fig.~\ref{fig4}, charge order and the LTT structures are strongly 
coupled, displaying that commensurability with the lattice is essentially important for 
stabilizing charge order. In this case, one can easily imagine that $\epsilon$ should 
have a commensurate value of 1/8, as predicted theoretically\cite{Vojta1999}. 
Tranquada {\it et al}. have noted that the incommensurate value of $\epsilon$ can be 
regarded as a {\em disordered} stripe in which there is the mixture of distinct stripe 
periods of $4a$ and $5a$\cite{Tranquada1999}. 
In scattering intensities calculated under this assumption, the charge order peak is 
broadened whereas the magnetic order peak remains sharp.
In fact, in our LBSCO system, the intrinsic line-width of the superlattice peak along 
$H$-direction is considerably broader than that of resolution-limited magnetic peaks 
observed by neutron scattering\cite{Fujita2002,Fujita2001}. 
These results imply that charge stripe order in cuprates is intrinsically disordered in 
comparison with that of isostructural systems of La$_{2-x}$Sr$_{x}$NiO$_{4}$ in 
which stripe order is mostly stabilized around commensurate positions with 
$\epsilon=1/3$\cite{Yoshizawa2000}. 
It should be noted that the high two-dimensionality of charge correlation 
($\xi_{a,b}/\xi_{c}>6$) could make the stripe correlation disordered. 

Line-broadening of the superlattice peak is seen along both the $H$-direction and the 
$K$-direction. In particular, these systematic experiments using single domain 
crystals have revealed that line-widths 
along the $K$-axis for orthorhombic $x=0.075$ crystals are much broader than for 
tetragonal $x=0.05$ crystals. Based on the stripe model, the line-width along $K$ 
corresponds to the mosaicity of the charge stripe. 
In addition, the orthorhombic symmetry of superlattice peaks in $x=0.075$ crystals 
suggests that the charge stripes are oblique.
As Fujita {\it et al}. have noted\cite{Fujita2002_2}, a corrugated pattern in the 
CuO$_{2}$ plane in LTLO phase can easily produce steps or kinks in the stripes, 
giving rise to the oblique of charge stripe. 
In this point of view, more oblique stripes could introduce the steps or kinks 
more randomly, which yields charge stripe mosaicity. 
Therefore, oblique stripe order becomes more disordered or {\em smectic} 
in comparison with the aligned stripe, consistent with the present results. 

In the LTT phase, the tilting pattern of the CuO$_{6}$ octahedra, i.e. the lattice potential 
pattern, is rotated by 90$^{\circ}$ with respect to the nearest-neighbor layers.
Thus the wave vector of charge order is rotated by 90$^{\circ}$.
Furthermore, the phase of charge order is shifted by $\pi$ from the n.n.n. layer to minimize 
the energy losses due to Coulomb interactions, giving rise to a twofold periodicity along 
the $c$ axis. Therefore, the 2$c$ periodicity of the superlattice peaks suggests that 
the doped holes are indeed arranged one-dimensionally across the two dimensional 
CuO$_{2}$ plane. 

%\section{Conclusion}
In conclusion, the propagation wave vector and three dimensional correlation of charge order 
in LBSCO systems were determined accurately using high-intensity synchrotron X-ray diffraction. 
Despite the 1/8-hole doping, the incommensurability of the superlattice peak 
($\epsilon =0.12$~r.l.u.) is clearly shifted away from the commensurate value of 1/8, indicating 
that charge stripe order in cuprates is intrinsically disordered. 
The orthorhombic $x=0.075$ crystal provided detailed information about the peak shift 
as well as the line width of the superlattice peak, indicating that the oblique stripes 
in $x=0.075$ crystal are more disordered than the aligned stripes in $x=0.05$ crystal.
The charge order was also found to be 2$c$ periodic and two-dimensional in nature. 
A proper determination of the atomic displacement pattern associated with the charge order 
is required to fully understand the essential nature of the (disordered) charge stripe order.
We thank K. Machida, M. Matsuda, H. Yamase, and J. M. Tranquada for valuable discussions.
This work was supported in part by a Grant-In-Aid for the Encouragement of Young 
Scientists (13740198 and 13740216, 2001), Scientific research B (14340105, 2000), 
Scientific Research on Priority Areas (12046239, 2001), and 
Creative Scientific Research (13NP0201) from the Japanese Ministry of Education, 
Science, Sports and Culture, and by the Core Research for Evolutional Science and 
Technology (CREST) from the Japan Science and Technology Corporation.
The synchrotron X-ray experiments were carried out at the SPring-8 facility with 
the approval of the Japan Synchrotron Radiation Research Institute 
(Proposal No. 1998A0157, 1999A0234, and 2000B0321). 
%\end{acknowledgments}

%\bibliography{references}

\end{document}